\documentclass[%
reprint,
 amsmath,amssymb,
 aps,
 prx,
]{revtex4-2}

\usepackage{graphicx}
\usepackage{subfig}
\usepackage{dcolumn}
\usepackage{bm}
\usepackage{hyperref}
\usepackage{ragged2e}
\usepackage{placeins}
\usepackage[version=4,arrows=pgf-filled,
textfontname=sffamily,
mathfontname=mathsf]{mhchem}
\usepackage{lipsum}

\begin{document}

\preprint{APS/123-QED}

\title{\textbf{Generation of Large Coherent-State Superpositions in Free-Space Optical Pulses}}
\author{Lucas Caron}
\author{Hector Simon}
\author{Hugo Basset}
\author{Romaric Journet}
\author{Rosa Tualle-Brouri}
\affiliation{
Laboratoire Charles Fabry, Institut d’Optique Graduate School, CNRS, Université Paris-Saclay,\\
2 Avenue Augustin Fresnel, 91127 Palaiseau, France}
\email{Contact author: rosa.tualle-brouri@institutoptique.fr}

\date{\today}
\begin{abstract}
The generation of non-Gaussian quantum states is a key requirement for universal continuous-variable quantum information processing.
We report the experimental generation of large-amplitude squeezed coherent-state superpositions (squeezed cat states) on free-space optical pulses, reaching an amplitude of $\alpha = 2.47$, which, to our knowledge, exceeds all previously reported values.
Our protocol relies on the controlled mixing of the Fock states $|1\rangle$ and $|2\rangle$ through a tunable beam splitter, followed by heralding via homodyne detection.
The resulting state displays three well-resolved negative regions in its Wigner function and achieves a fidelity of $0.53$ with the target state $\propto \hat{S}(z)(|\alpha\rangle - |-\alpha\rangle)$, with $\alpha = 2.47$ and squeezing parameter $z = 0.56$.
These results constitute a significant milestone for temporal breeding protocols and for the iterative generation of optical GKP states, opening new perspectives for scalable and fault-tolerant photonic quantum architectures.
\end{abstract}
\maketitle
\section{Introduction}
\label{section:intro}

Photonic platforms are promising candidates for the development of quantum computing. In principle, they enable room-temperature operation while remaining modular and naturally compatible with quantum networks~\cite{bourassa_blueprint_2021}. Significant progress has recently been made toward scalable architectures~\cite{aghaee_rad_scaling_2025}. 
However, encoding qubits in Gottesman–Kitaev–Preskill (GKP) states~\cite{gottesman_encoding_2001}, which allow universal and fault-tolerant quantum computation using only Gaussian operations, is a key milestone to fully unlock the potential of these architectures.

Recent experiments have achieved important milestones in the on-chip generation of such states~\cite{larsen_integrated_2025}. In parallel, extensive efforts on free-space platforms have enabled the production of increasingly complex non-Gaussian states, typically squeezed cat states, and demonstrated the first iterations of temporal breeding protocols~\cite{sychev_enlargement_2017, konno_logical_2024, simon_experimental_2024, hanamura_scalable_2025}, laying the groundwork for free-space GKP state generation.

In this article, we report what is, to our knowledge~\cite{gerrits_generation_2010, sychev_enlargement_2017, konno_logical_2024, neergaard-nielsen_generation_2006, huang_optical_2015, hacker_deterministic_2019, etesse_experimental_2015, simon_experimental_2024, hanamura_scalable_2025}, the first generation of free-space optical coherent-state superpositions with $\alpha > 1.9$, reaching an amplitude of $\alpha = 2.47$. Combined with the temporal multiplexing capabilities of our architecture~\cite{simon_experimental_2024}, this highly non-Gaussian resource brings us closer to the realization of fully fledged optical GKP states.

\begin{figure}[h!]
    \centering
    \includegraphics[width=\linewidth]{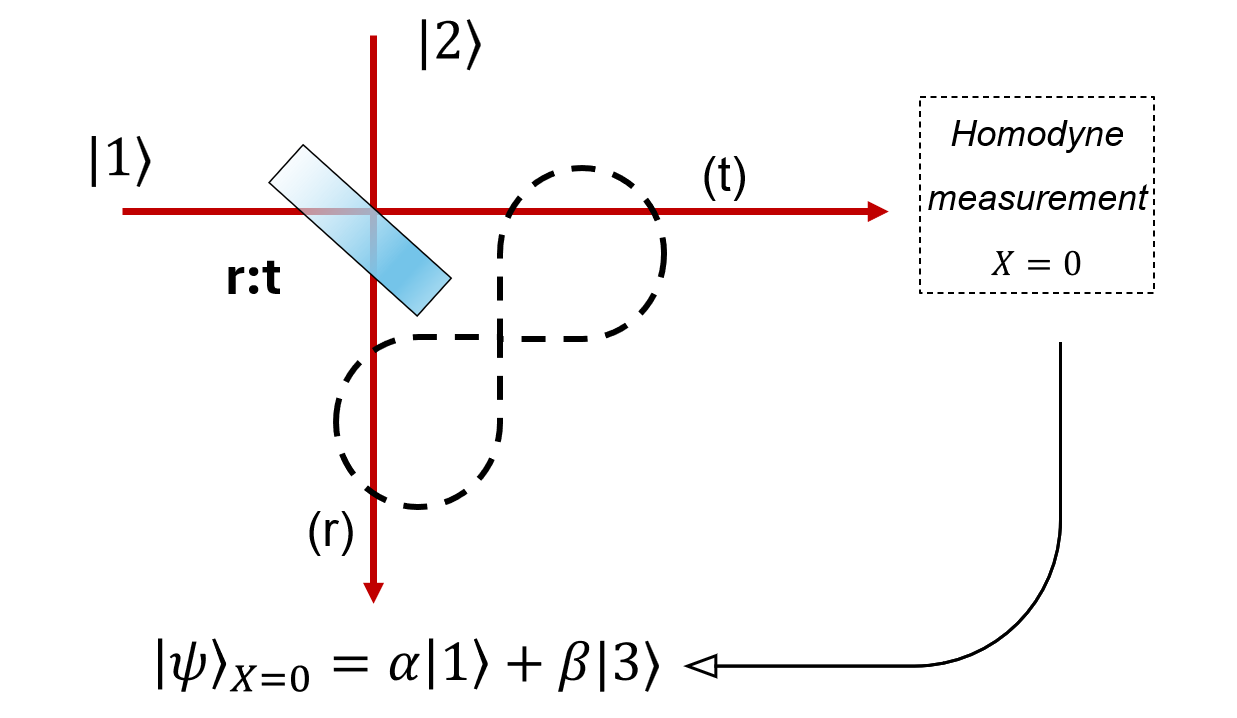}
    \caption{\justifying Fock states $|1\rangle$ and $|2\rangle$ enter the two input ports of a beam splitter with amplitude reflectivity~$r$ and transmissivity~$t$, and a homodyne measurement is performed on one output port.}
    \label{fig:bs}
\end{figure}

\begin{figure}[h]
    \centering
    \includegraphics[width=0.9\linewidth]{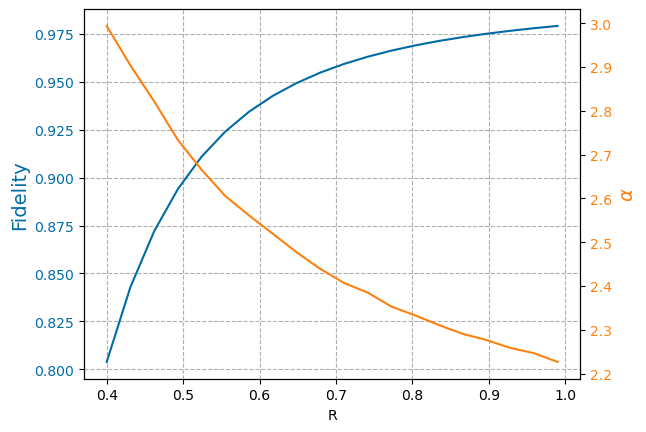}
    \caption{\justifying Fidelity between the state $|\psi\rangle_{X=0}$ and the closest SCSS $\hat{S}(z)(|\alpha\rangle - |-\alpha\rangle)$ as a function of $R=r^2$ (blue). The corresponding cat-state amplitude $\alpha$ is shown in orange.}
    \label{fig:ideal_cat}
\end{figure}

\indent We consider coherent-state superpositions (CSS), also referred to as odd cat states, defined as
\begin{equation}
    |CSS(\alpha)\rangle = \mathcal{N}_\alpha \left( |\alpha\rangle - |-\alpha\rangle \right),
\end{equation}
where $\mathcal{N}_\alpha$ is a normalization factor. In our experiment, we generate squeezed coherent-state superpositions (SCSS) obtained by applying the squeezing operator
\[
\hat{S}(z) = e^{\frac{1}{2}(z^{*}\hat{a}^2 - z\hat{a}^{\dagger 2})}.
\]

\FloatBarrier
The core of the protocol is the controlled mixing of the Fock states $|1\rangle$ and $|2\rangle$ on a tunable beam splitter, followed by a heralding homodyne measurement on one of the output modes.

\begin{figure}[t!]
    \centering
    \includegraphics[width=\linewidth]{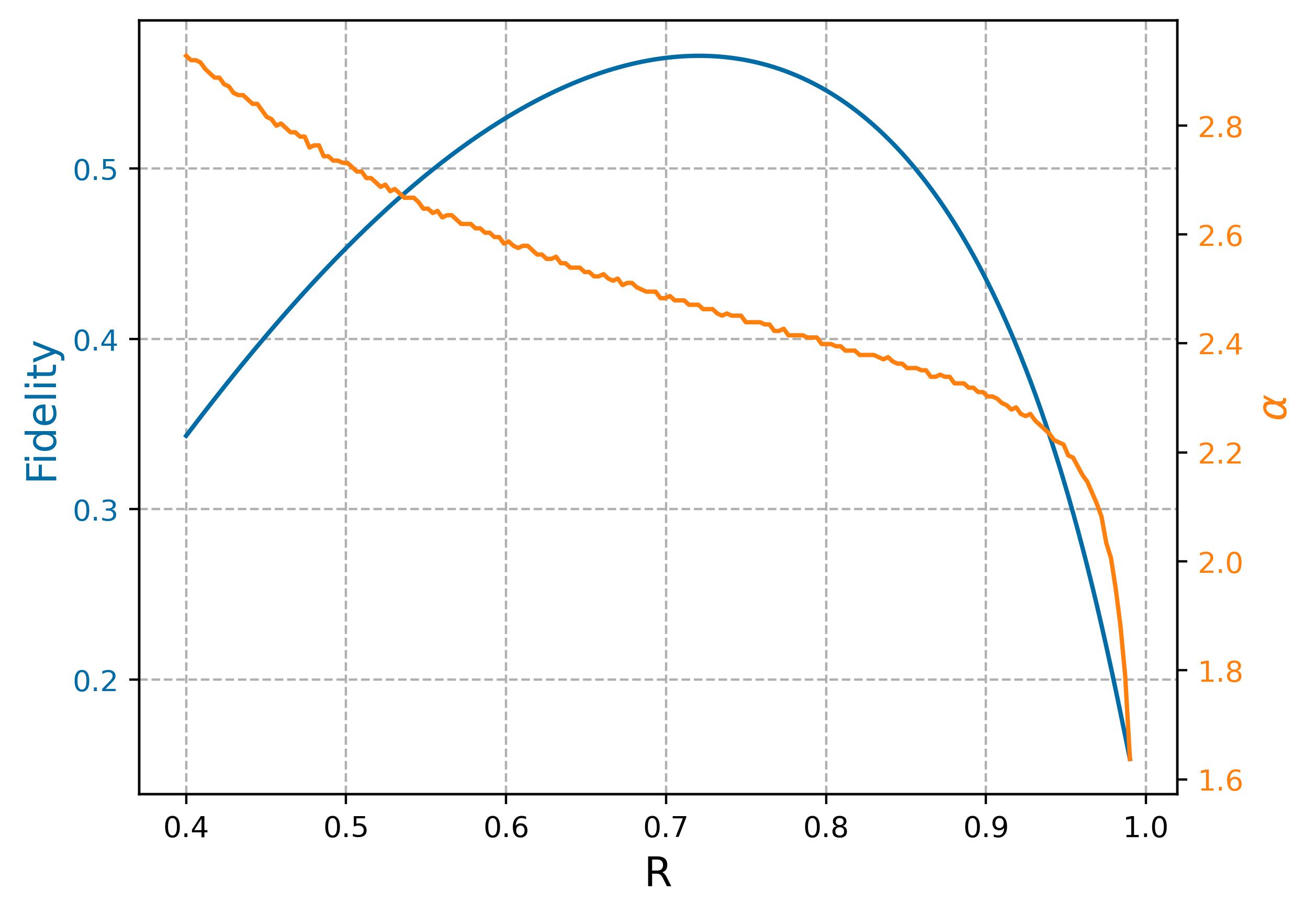}
    \caption{\justifying Simulated fidelity with the closest SCSS $\hat{S}(z)(|\alpha\rangle - |-\alpha\rangle)$ as a function of $R=r^2$ (blue), and corresponding amplitude $\alpha$ (orange).}
    \label{fig:fidelity_exp}
\end{figure}

Let us call $|\psi\rangle_{X=0}$ the state after the heralding homodyne measurement X = 0 (with $\hat{X}=(\hat{a}+\hat{a}^\dagger)/\sqrt{2}$), and r the reflectivity of the beam splitter. One can show that the situation depicted in Figure \ref{fig:bs} leads to:
\[
|\psi\rangle_{X=0} \propto (1-3r^2)|1\rangle - \sqrt{6}\, r^2 |3\rangle .
\]

Depending on the value of $r$, this superposition approximates a SCSS. Figure~\ref{fig:ideal_cat} shows the fidelity \footnote{Defined as $F(\rho_1,\rho_2)=\left[\mathrm{tr}\!\left(\sqrt{\sqrt{\rho_1}\rho_2\sqrt{\rho_1}}\right)\right]^2$.} between $|\psi\rangle_{X=0}$ and the closest SCSS, obtained by optimizing over both the squeezing parameter $z \in \mathbb{R}$ and the amplitude $\alpha \in \mathbb{R}$. The corresponding optimal amplitude is also displayed.

In practice, the experimentally generated state deviates from the ideal superposition because of optical losses and the finite efficiency of the homodyne detector, among other factors. A realistic model of the setup, described in Appendix~\ref{appendixA}, allows us to simulate the expected output state under these conditions. When these imperfections are included, the dependence of the fidelity on the beam splitter reflectivity differs from that predicted by the idealized model of Fig.~\ref{fig:ideal_cat}. As shown in Fig.~\ref{fig:fidelity_exp}, the realistic model exhibits a maximum fidelity around a reflectivity of $R \simeq 0.72$. At this value, the predicted state reaches a fidelity of $0.57$ with an SCSS of amplitude $\alpha \simeq 2.47$ and squeezing of $4.82\,\mathrm{dB}$ (corresponding to $z \simeq 0.56$).

\section{Experimental setup} \label{section:exp}
\begin{figure}[h]
    \centering
    \includegraphics[width=\linewidth]{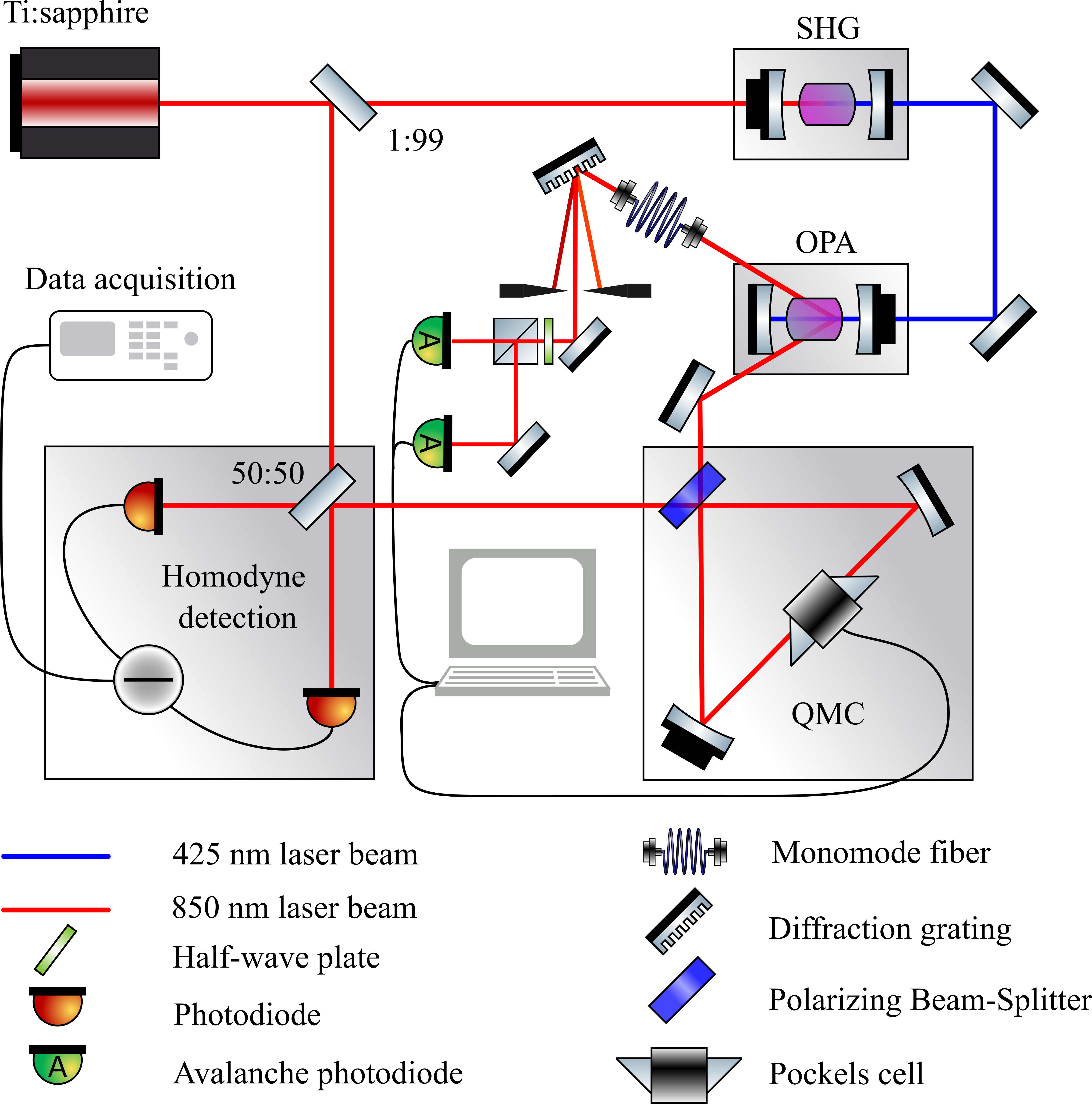}
    \caption{\justifying Simplified schematic of the experimental setup, showing the main building-blocks: the resource state source, the QMC and the homodyne detection.}
    \label{fig:exp_setup}
\end{figure}

A simplified schematic of the experimental setup is shown in Fig.~\ref{fig:exp_setup}. A mode-locked Ti:Sapphire laser produces 2.2~ps pulses at 850~nm with a repetition rate of 76~MHz. The resource states are generated through two nonlinear processes. First, a cavity-enhanced second-harmonic generation stage based on a \ce{BiB3O6} crystal produces a blue pump beam at 425~nm~\cite{kanseri_efficient_2016}. This beam subsequently pumps an optical parametric amplifier (OPA) in a resonant cavity, with an intracavity average power reaching 6~W. The OPA contains a \ce{BiB3O6} crystal in a noncollinear type-I phase-matching configuration and generates a two-mode squeezed vacuum (TMSV) state.

One mode of the TMSV is used to herald the presence of photons in the other mode. This arm is first spatially filtered by coupling it to a single-mode fiber, with a transmission exceeding 85\% (insertion loss $<0.7$~dB). The beam is then spectrally filtered using a diffraction grating, where the $-1$ diffraction order then goes toward a fine-tunable slit. After this spatial and spectral filtering stage, the beam passes through a half-wave plate and a polarizing beam splitter, whose two outputs are directed onto avalanche photodiodes (APDs). A detection event on one APD heralds a single-photon Fock state in the other mode of the TMSV, while simultaneous clicks on both APDs herald a two-photon Fock state. 

In the experiment, single-photon detection events occur at a typical rate of 400~kHz, while double detections occur at about 1~kHz. The electronic signals from the two APDs are summed, and a trigger threshold is applied to identify coincidence events. The other mode of the TMSV is sent through a 60~m-long Herriott-cell-like delay line, introducing a 200~ns delay between a detection event and the arrival of the resource state at the quantum memory cavity (QMC). This delay ensures that the electronics have enough time to process the detection event and trigger the storage sequence. In this way, a heralded single-photon state can be stored in the QMC while waiting for the heralding of a two-photon state.

The QMC is a low-loss optical cavity built from high-reflectivity mirrors ($>99.99\%$ in $s$-polarization), a polarizing beam splitter (PBS) at the entrance ($R_s > 99.95\%$ and $R_p <2\%$), and a fast Pockels cell (PC). The dominant losses originate from the PC; measurements indicate a round trip loss of around 1\% (see Appendix~\ref{appendix_losses}). The PC acts as a voltage-controlled wave plate placed just before the PBS. As a result, the \{PC + PBS\} assembly behaves as a tunable beam splitter whose effective reflectivity is set by the retardance induced by the PC (see Appendix~\ref{appendixA}). A detailed description of the QMC can be found in Refs.~\cite{simon_experimental_2024, cotte_experimental_2022, bouillard_quantum_2019}.
 
\begin{figure}[h!]
    \centering
    \includegraphics[width=1\linewidth]{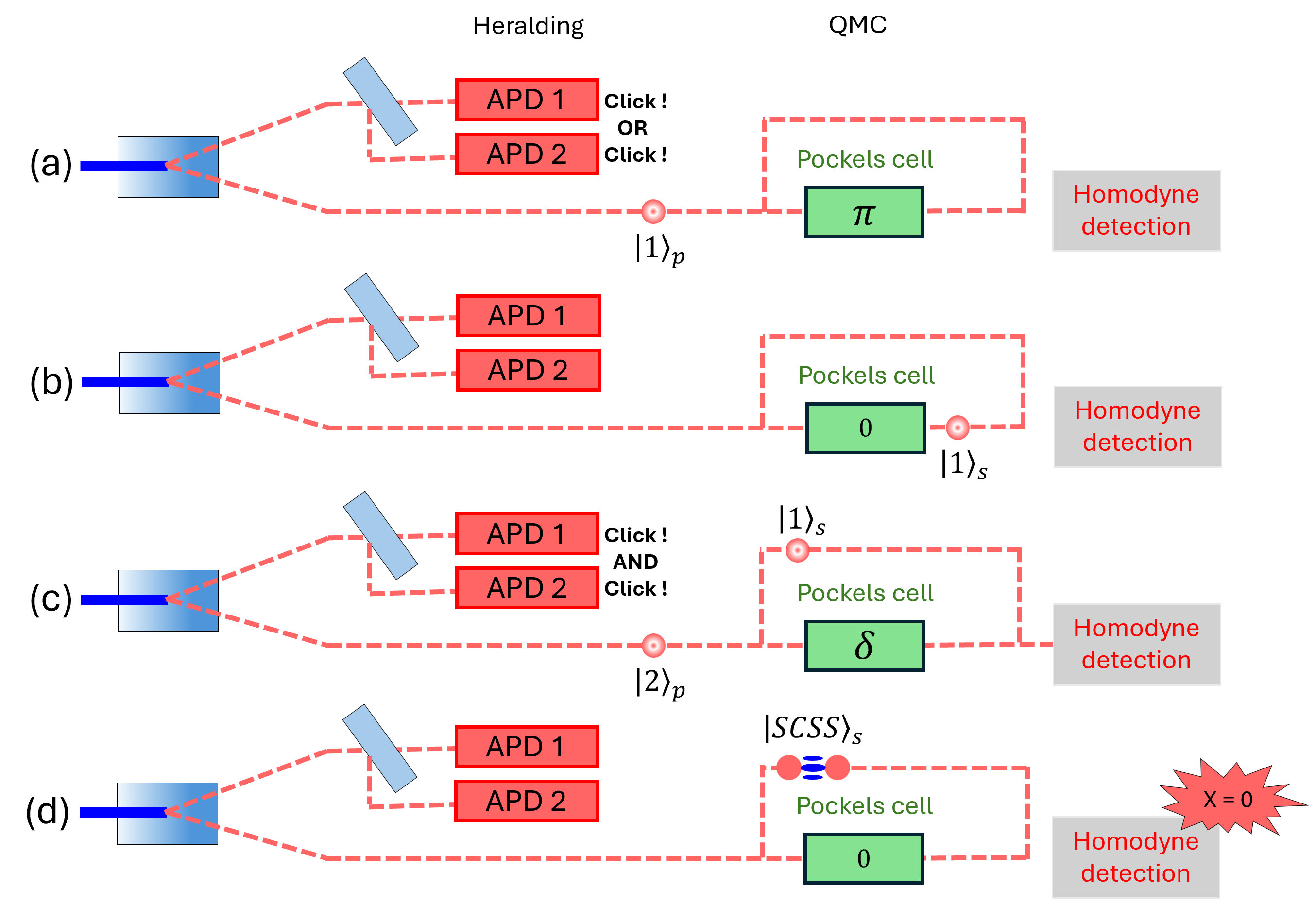}
    \caption{\justifying Diagram of the different steps of one experimental cycle : (a) heralding of a $p$-polarized single photon, half-wave plate operation by the PC. (b) The single photon is stored inside the QMC. (c) A $p$-polarized $|2\rangle_p$ is heralded and the PC apply a $\delta$ retardance to the $|2\rangle_p|1\rangle_s$ state. (d) A quadrature measurement is performed on the output path. If the result is $X=0$, the state in the QMC is the targeted state.}
    \label{fig:protocole}
\end{figure}

The resource states enter the QMC with $p$-polarization. Storage and on-demand release are implemented by applying an effective half-wave plate operation using the Pockels cell. In the experiment, we first store a single-photon state while waiting for a two-photon Fock state to be heralded, as illustrated in Fig.~\ref{fig:protocole}(a)--(b).

When the $|2\rangle$ state arrives, the two resource states carry orthogonal polarizations ($s$ and $p$), as shown in Fig.~\ref{fig:protocole}(c). By applying an arbitrary wave plate operation with the PC, the situation becomes equivalent to the beam splitter interaction of Fig.~\ref{fig:bs}. The optimal reflectivity $R \simeq 0.72$ obtained from Fig.~\ref{fig:fidelity_exp} corresponds to a PC-induced retardance of $\delta \simeq 2.03$~rad.

The round trip time of the QMC matches the laser repetition rate ($\sim$13.1~ns, corresponding to $\sim$3.9~m), which requires a Pockels cell with a rise time well below 13~ns. At the output of the QMC, a homodyne detector with 76\% efficiency is used both for the heralding measurement at $X=0$ [Fig.~\ref{fig:protocole}(d)] and for the characterization of the generated state. The homodyne bandwidth is about 2~MHz; to mitigate crosstalk with the heralding measurement, the produced state is stored in the QMC for 195~ns before extraction (15 round trips).

Finally, since the relative phase between the local oscillator and the signal is not actively stabilized, a phase measurement is required. Following the method of Ref.~\cite{cotte_experimental_2022}, we probe the phase drift accumulated during storage by successively sending two coherent states into the QMC. These states are stored for 1 and 15 round trips, respectively, and measured with the homodyne detector. Subtracting the two measurements gives access to the relative phase between the local oscillator and the produced state. All electronic control is performed using BME Bergmann delay generators, and data acquisition is carried out with a fast oscilloscope.

\section{Results}

The generated states are reconstructed using quantum state tomography (QST) \cite{lvovsky_continuous-variable_2009}. For each produced state, we measure the quadrature $X_\theta$ with the homodyne detection, where $\theta$ is the relative phase between the local oscillator and the state. This phase fluctuates freely and we measure it after each SCSS generation. Consequently, the tomography is done on a uniformly distributed phase set. The QST is performed using a maximum-likelihood algorithm~\cite{lvovsky_iterative_2004} on a dataset of 16{,}339 quadrature measurements. We compute the fidelity between the reconstructed density matrix and the squeezed coherent-state superposition $\hat{S}(z)(|\alpha\rangle - |-\alpha\rangle)$ with $\alpha=2.47$ and $z = 0.56$ (which is the SCSS closest to the simulated state expected from our protocol).
\begin{figure*}[t!]
    \centering
    \subfloat[]{\includegraphics[width=0.4\textwidth]{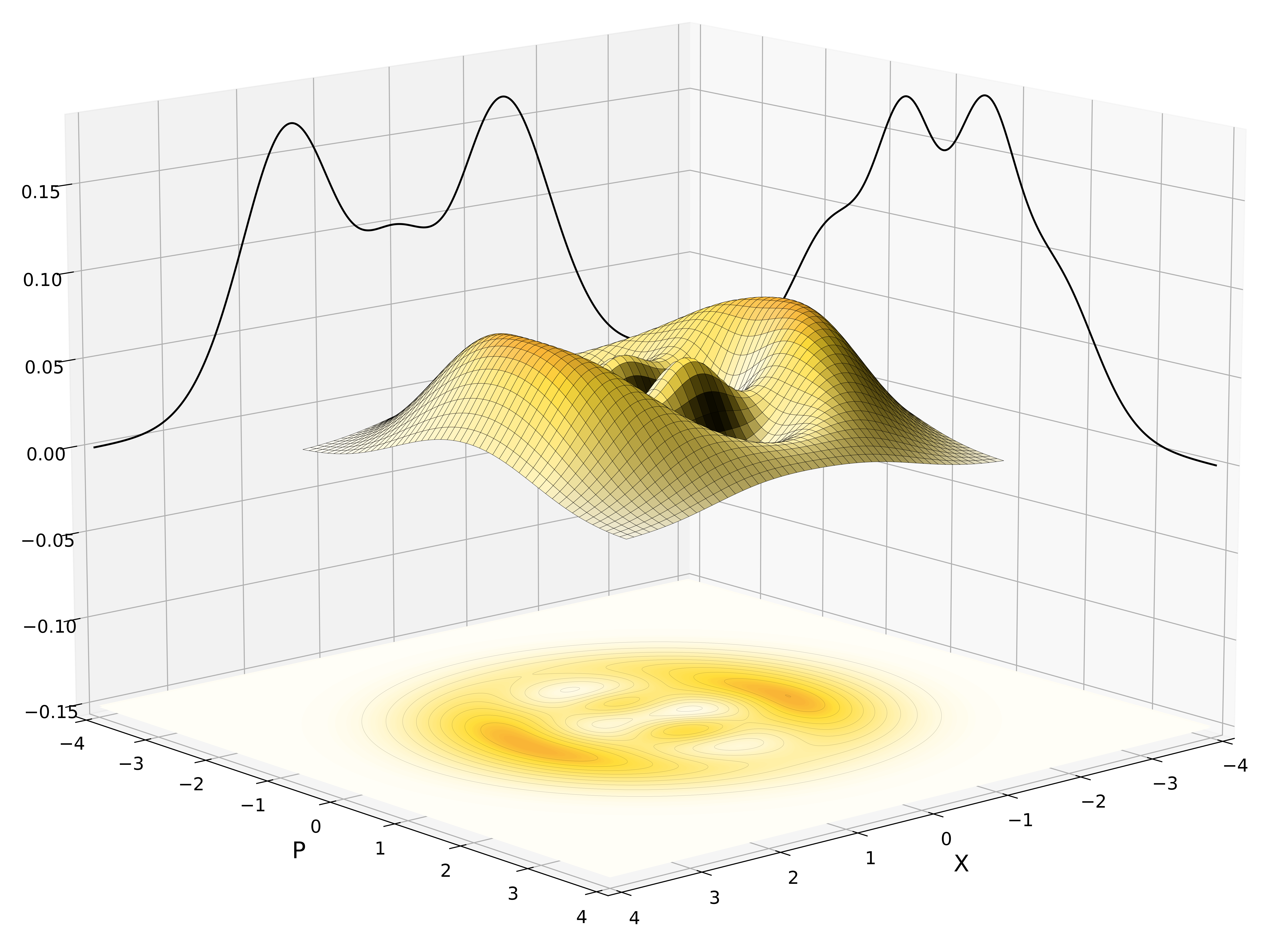}} 
    \subfloat[]{\includegraphics[width=0.4\textwidth]{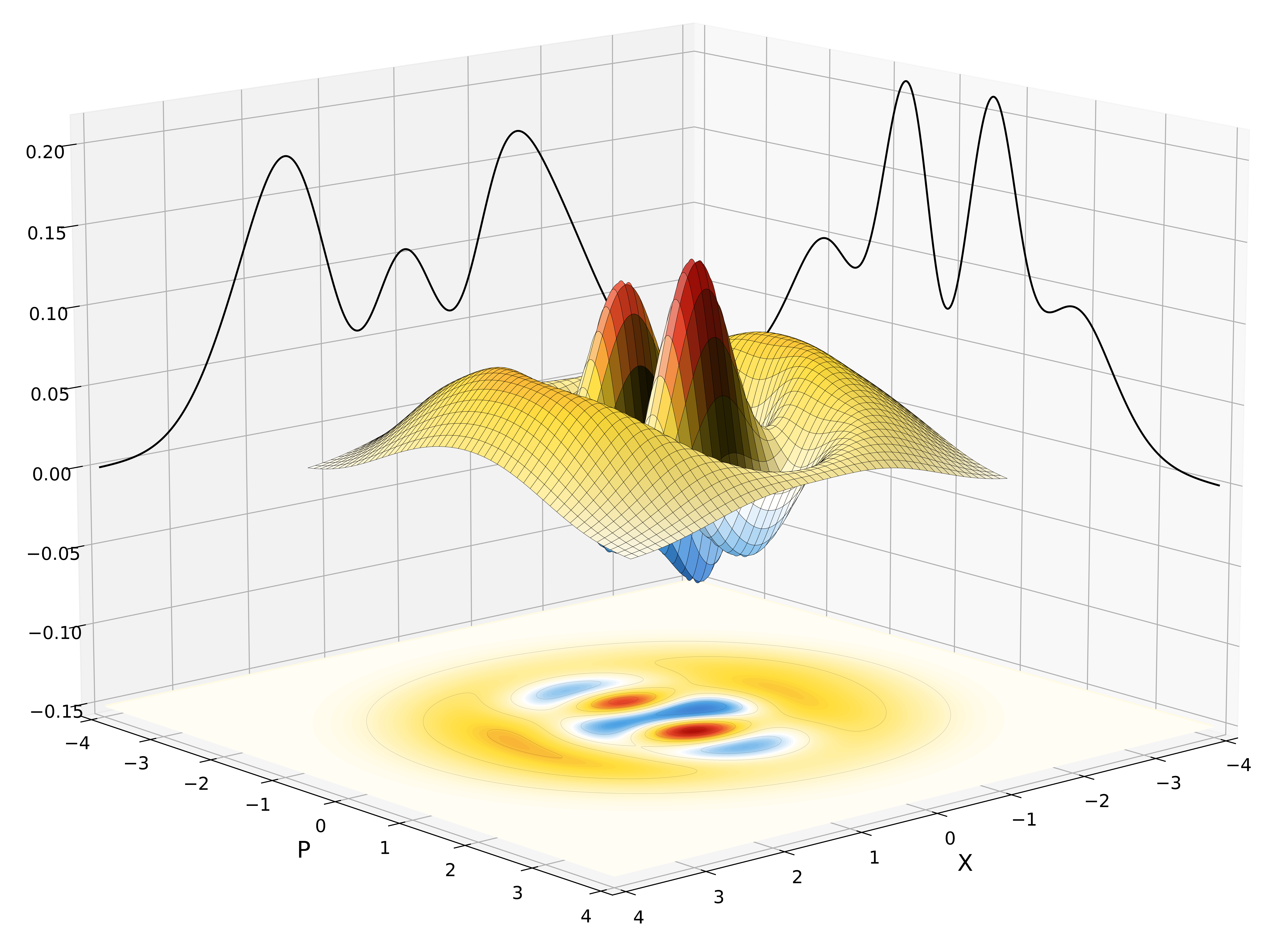}}
    \\
    \subfloat[]{\includegraphics[width=0.5\textwidth]{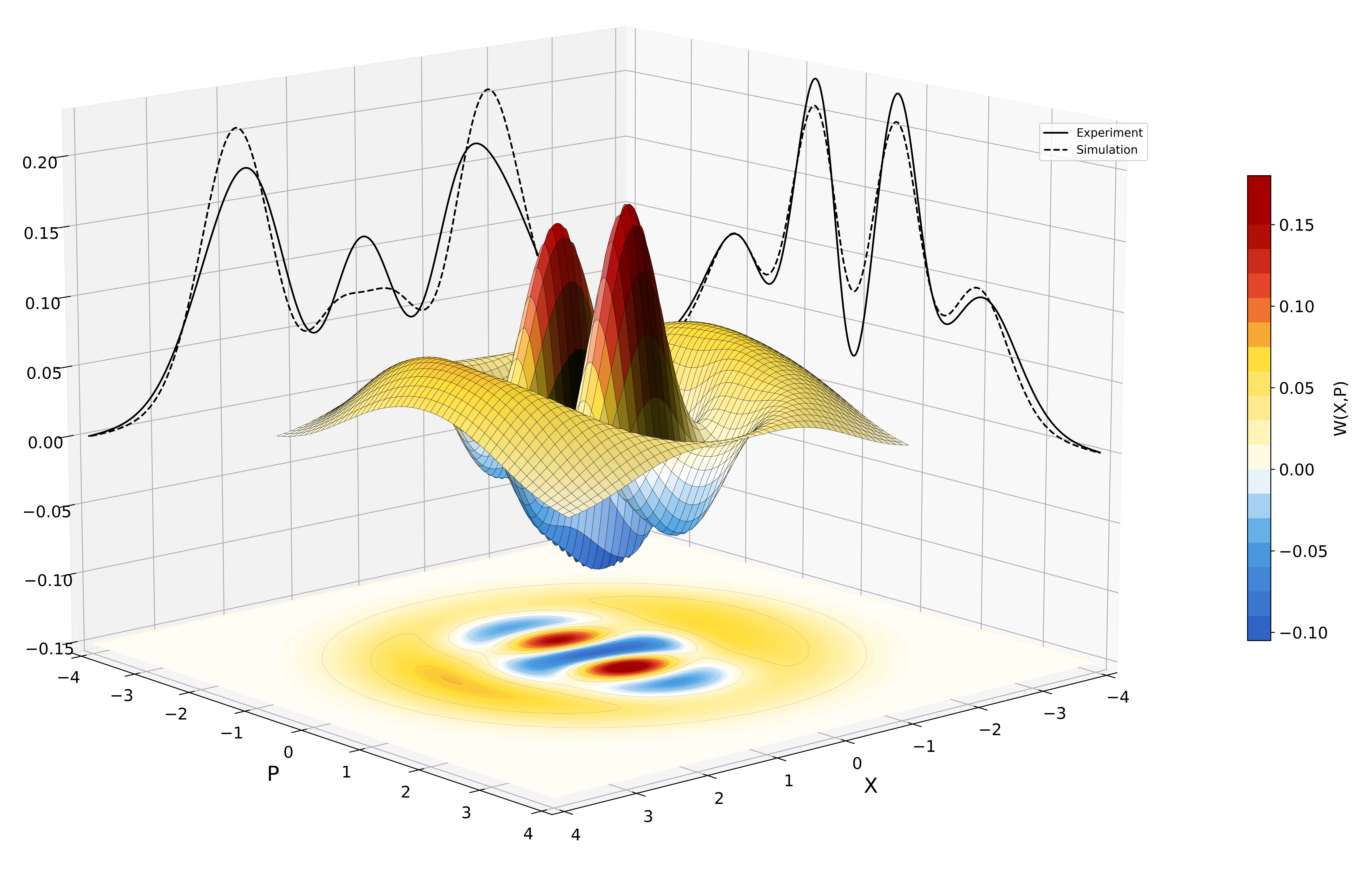}}
    \caption{\justifying Wigner functions reconstructed by QST (a) without correction, (b) corrected from homodyne detection efficiency, (c) corrected from homodyne detection efficiency and storage losses. The dotted lines on the marginal distributions represent the simulated state accounting for all the experimental losses.}
    \label{fig:res}
\end{figure*}

Figure~\ref{fig:res} shows the states obtained using a heralding window $X \in [-0.2,\,0.2]$ and by allowing the first state ($|1\rangle$) to be stored for up to 18 round trips ($\simeq 234$~ns). The first Wigner function [Fig.~\ref{fig:res}(a)] corresponds to the QST performed on the raw data, without applying any correction. This state is not the one actually generated inside the QMC, as the dominant losses arise from the limited efficiency of the homodyne detector. In Fig.~\ref{fig:res}(b), we correct only for the homodyne detection efficiency. The reconstructed state, which has undergone 15 round trips of storage in the QMC, still exhibits three well-resolved negative regions in its Wigner function. Finally, Fig.~\ref{fig:res}(c) shows the state that was actually produced immediately after the heralding measurement— and that would be immediately available for future use. We recall that storage is required solely because the same homodyne detector is used for both heralding and tomography, and its limited bandwidth forces the state to remain in the QMC during 195~ns before extraction. In Fig.~\ref{fig:res}(c), the reconstructed state is corrected for both the 76\% homodyne efficiency and the storage losses inside the QMC ($1 - 0.99^{15} \simeq 14\%$). This corrected state closely matches the simulated one (dotted curves on the marginals), with a fidelity of 93\%. The residual discrepancies mainly originate from slow drifts of the experimental parameters over the acquisition time (such as small variations in the laser repetition rate or power).

The fidelity between this corrected state and the ideal SCSS defined in Section~\ref{section:intro} is
\[
F = 0.53^{+0.01}_{-0.06}.
\]
The statistical uncertainty on the fidelity is obtained via a parametric-bootstrap resampling procedure~\cite{efron_introduction_1994} using 100 repetitions.
The corresponding generation rate is approximately 3.1~Hz. Higher rates can be achieved by allowing the first photon to be stored longer inside the QMC, at the expense of state quality. For example, when increasing the allowed storage time to a maximum of 24 round trips, we obtain 22\,695 data points, a fidelity of 0.51, and a generation rate of 4.25~Hz.

Furthermore, we can also point out that our protocol also produces, in most cases, even cat-states when a single-photon state is heralded rather than a two-photon Fock state. The simulated state has a fidelity of 0.61 with an even cat-state of amplitude $\alpha = 1.71$ and squeezing of 3.90 dB. We present, in the Appendix~\ref{appendix_even_cat}, a typical experimental result of even cat-state obtained in the same acquisitions as those of odd cat-state of Figure \ref{fig:res}.

\FloatBarrier
\section{Conclusion}

In this work, we have demonstrated the generation of free-propagating optical odd SCSS (cat) states with an unprecedented amplitude of $\alpha = 2.47$. Notably, the reconstructed Wigner functions display three well-resolved negative regions even after a 195~ns storage time in the quantum memory cavity. By leveraging temporal multiplexing, the experiment implements iterative breeding operations. All components required for a two-step large-SCSS breeding protocol have now been demonstrated, paving the way toward optical GKP state generation.

Several technical improvements can substantially increase the generation rate. Replacing the heralding APDs with high-efficiency superconducting nanowire single-photon detectors (SNSPDs) would provide a significant boost. In addition, the current protocol relies on post-processing, and real-time analysis would further enhance the achievable rates. Finally, adding additional quantum memory cavities would enable more complex iterative protocols, opening a clear path toward the generation of higher-complexity non-Gaussian states in the near future.

\begin{acknowledgments}
This work was supported by the Agence Nationale de la Recherche with the IGNITION project (Grant No. ANR-21-CE47-0015-01), the NISQ2LSQ PEPR project (Grant No. ANR-22-PETQ-0006), OQULUS PEPR project (Grant No. ANR-22-PETQ-0001), and a PhD funding from QuanTEdu-France (Grant No. ANR-22-CMAS-0001).
\end{acknowledgments}


\appendix
\section{Model of the experiment \label{appendixA}}
\subsection{$|1\rangle$ and $|2\rangle$ on a r:t beam splitter}
\noindent We start from a two-mode squeezed vacuum state: $$|\text{TMSV}\rangle = \frac{1}{\cosh{r}}\sum_{n=0}^\infty(e^{-i\phi}\tanh{r})^n|nn\rangle$$      
When an event is detected on one of the APDs, the state is approximately described, for $r \ll1$, by the following density matrix:

$$\hat{\rho}_1\propto |1\rangle\langle1| + r^2(2-\eta)|2\rangle\langle2|$$  
where $\eta$ is the overall efficiency of the heralding
channel, modeled with a beam splitter of transmission $\eta$. 
We estimate $\eta\approx 0.15$ in the experiment, and one can deduce the value of $r$ from:
    
$$\frac{\text{APD click rate}}{\eta f_\text{pump}} = r^2$$    
with $f_\text{pump} = 76$ MHz. Similarly, the state that is heralded when both APDs click at the same time can be written:
$$\hat{\rho}_2\propto p_2r^4|2\rangle\langle2| + p_3r^6|3\rangle\langle3|$$    
where $p_2$ (resp. $p_3$) are the probabilities of obtaining a click on each APD, knowing that a state $|2\rangle$ (resp. $|3\rangle$) has been generated by the OPA in the mode of
the fiber and the monochromator.\\    
For $p_2$: both photons are transmitted during the filtering stage, then after the beam splitter they exit in both modes and are detected by each of the APDs.
$$p_2 = \frac{\eta^2}{2}$$\\
For $p_3$: let $\eta_f$ be the filtering efficiency and $\eta_\text{APD}$ the efficiency of the APDs (with $\eta = \eta_f \eta_\text{APD}$). In a first scenario, the three photons are transmitted (probability of $\eta_f^3$). On the one hand, the probability that the photons do not all exit in the same mode after the beam splitter is $\frac{3}{4}$ and on the other hand, the probability of having a click on each of the APDs is $\eta_\text{APD} \times \left[1 - (1-\eta_\text{APD})^2\right]$.
In a second scenario, only two photons pass through the filter (probability $3\eta_f^2(1-\eta_f)$) and the probability that both APDs click is $\frac{1}{2} \times \eta_\text{APD}^2$, so that:
    
$$p_3 = \frac{3}{4}\eta_f^3\eta_\text{APD}(1-(1-\eta_\text{APD})^2) + \frac{3}{2}\eta_f^2(1-\eta_f)\eta_\text{APD}^2$$
Finally:
$$p_3=\frac{3}{2}\eta^2\left(1-\frac{\eta}{2}\right)$$
\\
\\
Now, let's consider the losses that these states will experience. 
The losses must take into account the sum $\eta_{prop}$ of all optical losses before the QMC, and the internal losses of the QMC: $\eta_{QMC}$ for each turn.
The $|1\rangle$ states will be stored for $N_{stor}$ QMC round trips, while the 2-photon state undergoes only one round trip. As a loss $\eta$ can be modeled with a beam splitter of reflectivity $R=\eta$, one has:
\begin{widetext}

$$
\left\{
\begin{array}{ll}
\hat{\rho}_1^{'} = Tr_{2}(\hat{U}_1(\hat{\rho}_1\otimes \hat{\rho}_\text{vac})\hat{U}_1^{\dagger}) \text{ with } \hat{U}_1 = e^{\arcsin(\sqrt{R_1})(\hat{a}^\dagger_1\hat{a}_2 - \hat{a}_1\hat{a}^\dagger_2)} \text{ and } R_1 = 1 - (1- \eta_{prop}) \times (1-\eta_{QMC})^{N_{stor}} \\
\hat{\rho}_2^{'} = Tr_{2}(\hat{U}_2(\hat{\rho}_2\otimes \hat{\rho}_\text{vac})\hat{U}_2^{\dagger})  \text{ with } \hat{U}_2 = e^{\arcsin(\sqrt{R_2})(\hat{a}^\dagger_1\hat{a}_2 - \hat{a}_1\hat{a}^\dagger_2)} \text{ and } R_2 = 1 - (1- \eta_{prop}) \times (1-\eta_{QMC})
\end{array}
\right.
$$
\end{widetext}
\noindent where $\hat{\rho}_{vac}$ stands for the vacuum, and where the $\hat{a}_i$ are the ladder operators associated to the beam splitter modes. These two states then interact as inputs for the two modes of a beam splitter (1 and 2) of arbitrary reflectivity $R$: 
\begin{equation}{\label{eq:a1}}
\hat{\rho}_{12} = \hat{U}_{12}(\hat{\rho}_1^{'}\otimes\hat{\rho}_2^{'}) \hat{U}^\dagger_{12} \text{ with } \hat{U}_{12} = e^{\arcsin(\sqrt{R})(\hat{a}^\dagger_1\hat{a}_2 - \hat{a}_1\hat{a}^\dagger_2)}
\end{equation}
Before performing the heralding measurement, losses are applied to the heralding channel: in order to account for the $76\%$ efficiency of the homodyne detection. This is again modeled using a beam splitter of reflectivity $R_{HD} = 0.24$, leading to the two-mode density matrix $\hat{\rho}_{12}^{'}$.\\
One can now apply the partial projector associated with the quadrature $X=0$ heralding measurement :

$$\hat{P}_{X=0} = \mathcal{I}\otimes \langle X=0| \text{ with } |X=0\rangle = \sum_{n=0}^N \Psi_n(X=0)|n\rangle$$
where $\Psi_n(x)$ is the wavefunction of Fock state $|n\rangle$ in $X$-quadrature basis, $\mathcal{I}$ the identity and $N$ the size of the basis ($N=20$ in the simulations).
Eventually, the state in the other mode, after the heralding measurement, is:
$$
\hat{\rho}_{SCS} = \hat{P}_{X=0}\hat{\rho}_{12}^{'}\hat{P}_{X=0}^{\dagger}
$$
up to a normalization factor.\\
For each value of R, one can easily compute the squeezed coherent state superposition that is closest to the produced state.

\subsection{Pockels cell + PBS as a tunable beam splitter}

In the experiment, the tunable beam splitter is realized with a fast Pockels cell and a polarizing beam splitter (PBS). The Pockels cell acts as a wave plate introducing an arbitrary retardance $\delta$ between the eigenaxes. This wave plate is oriented at 45° from the horizontal plane.
Let us denote $\hat{a}_H$ and $\hat{a}_V$ the ladder operators associated with the horizontal and vertical polarization. The unitary operation of the Pockels cell of retardance $\delta$ writes:\\
\begin{equation}
    \hat{U}(\delta) = \exp{\left(i\frac{\delta}{2}(\hat{a}_H^\dagger \hat{a}_V + \hat{a}_V^\dagger \hat{a}_H)\right)}
\end{equation}

In particular, half-wave plate and quarter-wave plate are obtained for $\delta = \pi$  and $\delta = \pi/2$.
This operation is equivalent to the action of a beam splitter (equation \ref{eq:a1}) up to a meaningless $\pi/2$ phase-shift on one of the output modes (which adds to the global phase-shift experienced by this mode), and with $R=\sin^2(\delta/2)$.

\subsection{Simulation}
\begin{figure}
    \centering
    \subfloat[]{{\includegraphics[width=.8\linewidth]{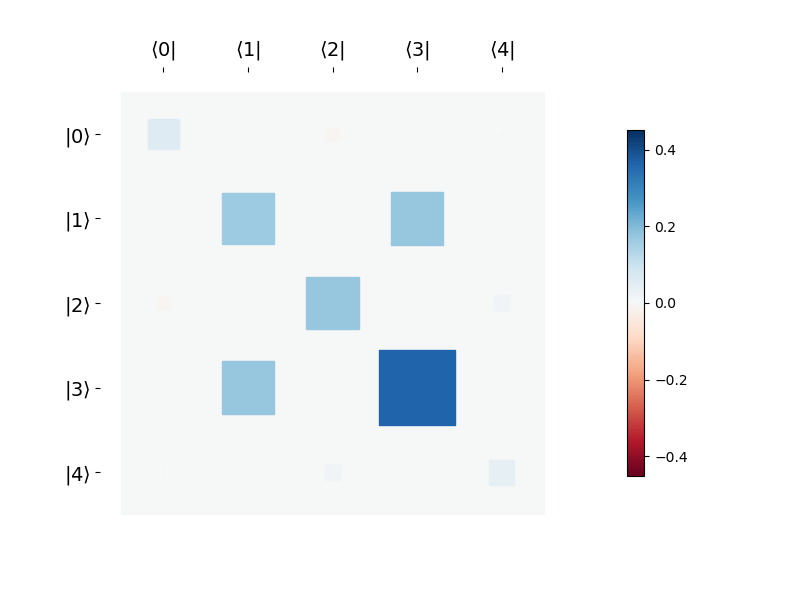} }}
    \\
    \subfloat[]{{\includegraphics[width=.8\linewidth]{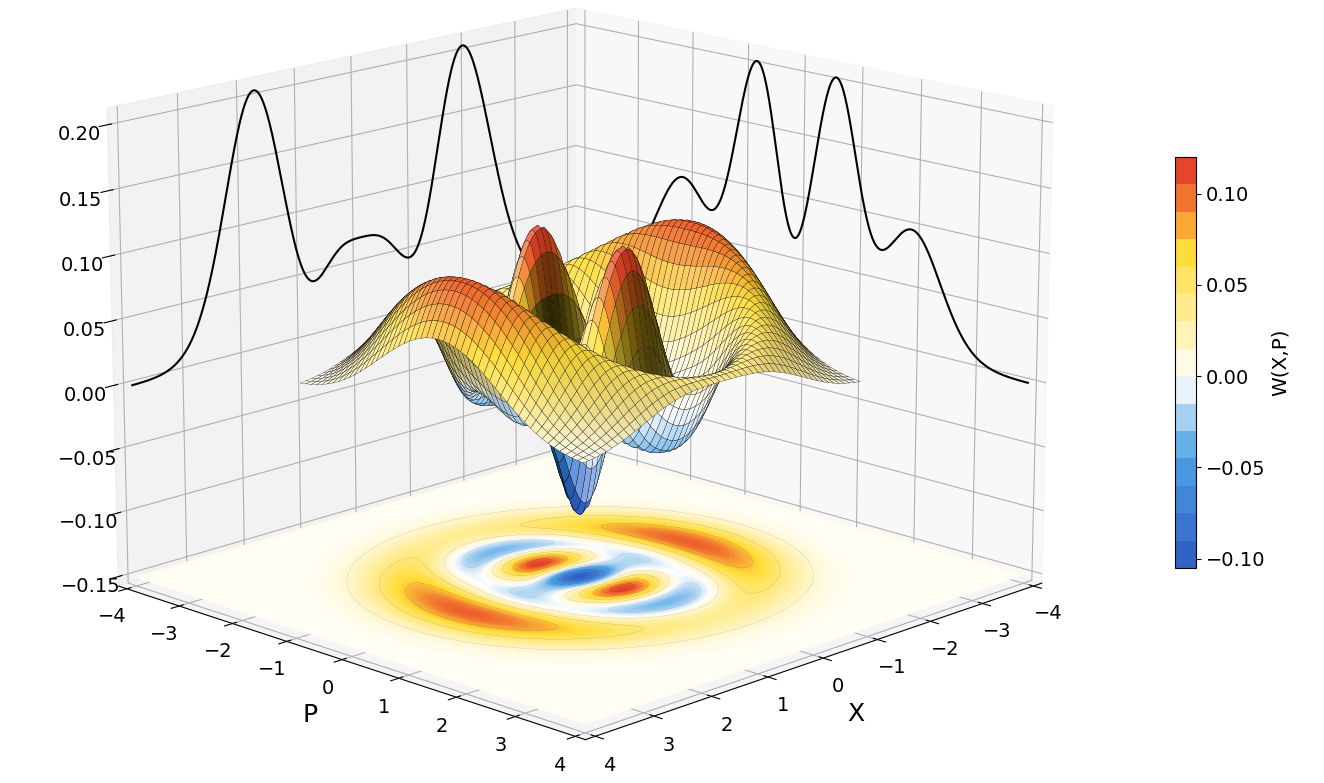} }}
    \caption{\justifying Simulation of the optimal state that can be generated by the experiment by storing the first single-photon during 14 round trips. (a) Hinton plot of the density matrix and (b) 3D Wigner function.}
    \label{fig:simulated_scss}
\end{figure}

Since the first photon is stored between 9 and 18 round trips in the experiment and since the distribution probability of the number of storage round trip is uniform, we realize the mean of the density matrices computed with $N_{stor} \in [9,18]$. We estimated the losses before the QMC at $\simeq 6\%$.
We find that the state is the closest to an SCSS for a retardance of 2.03 rad (equivalent to $R=0.72$). The state therefore has a fidelity of 0.57 with SCSS of amplitude $\alpha = 2.47$ with a squeezing of 4.82 dB (z = 0.56). 
This optimal state is depicted on Fig. \ref{fig:simulated_scss}.

\subsection{Experimental results: density matrices}
Table \ref{tab:dm} presents the density matrices associated to the Wigner function depicted in Figure~6 of the main paper.
\begin{table*}[t!]
\centering

\begin{tabular}{c c}
\hline\hline
$\rho_{SCS}\text{(a)}$\\
\hline
\\
\scriptsize{
$\begin{pmatrix}
0.23 & -0.01j & 0.06 & 0 & -0.02 & 0\\
0.01j & 0.31 & 0    & 0.09+0.01j & -0.01j & 0\\
0.06 & 0    & 0.27 & 0.01 & -0.02-0.01j & 0\\
0 & 0.09-0.01j    & 0.01    & 0.15 & 0 & 0.01 \\
-0.02 & 0.01j    & -0.02+0.01j    & 0 & 0.02 & 0 \\
0 & 0    & 0   & 0.01 & 0 & 0.01
\end{pmatrix}$
}
\\\\
\hline\hline $\rho_{SCS}\text{(b)}$ \\
\hline
\\
\scriptsize{
$\begin{pmatrix}
0.15 & -0.01j & 0.01-0.01j & 0.01+0.01j & -0.03 & 0.01j\\
0.01j & 0.25 & -0.01+0.01j    & 0.19+0.03j & -0.01-0.02j & -0.01j\\
0.01+0.01j & -0.01-0.01j & 0.24 & 0.02 & -0.07-0.02j & 0.01j\\
0.01-0.01j & 0.19-0.03j  & 0.02    & 0.28 & -0.01 & 0.04+0.01j \\
-0.03 & -0.01+0.02j    & -0.07+0.02j    & -0.01 & 0.04 & -0.01 \\
-0.01j & 0.01j    & -0.01j   & 0.04-0.01j & -0.01 & 0.04
\end{pmatrix}$
}
\\\\
\hline\hline
\multicolumn{2}{c}{$\rho_{SCS}\text{(c)}$} \\
\hline
\\
\multicolumn{2}{c}{

\scriptsize{
$\begin{pmatrix}
0.12 & -0.01j & -0.04-0.02j & 0.01j & -0.03 & 0.01j\\
0.01j & 0.21 & -0.01+0.01j    & 0.24+0.04j & -0.02j & -0.01j\\
-0.04+0.02j & -0.01-0.01j    & 0.17 & 0.02 & -0.07-0.02j & 0.01j\\
-0.01j & 0.24-0.04j    & 0.02    & 0.37 & 0 & 0.03 \\
-0.03 & 0.02j    & -0.07+0.02j    & 0 & 0.07 & -0.01+0.01j \\
-0.01j & 0.01j & -0.01j   & 0.03 & -0.01+0.01j & 0.06
\end{pmatrix}$
}
}

\\\\
\hline\hline
\end{tabular}
\caption{\justifying \small{Density matrix, in the Fock basis, of the states presented in Figure~6 of the
main text. The three cells represent the density matrix of: (a) the uncorrected SCS, (b) the SCS corrected for 76~\% detection efficiency and (c) the SCS corrected for 76~\% detection efficiency and 15 round trip storage.}}
\label{tab:dm}

\end{table*}

\subsection{Even cat-states}
\subsubsection{Simulation}
\begin{figure}
{{\includegraphics[width=.8\linewidth]{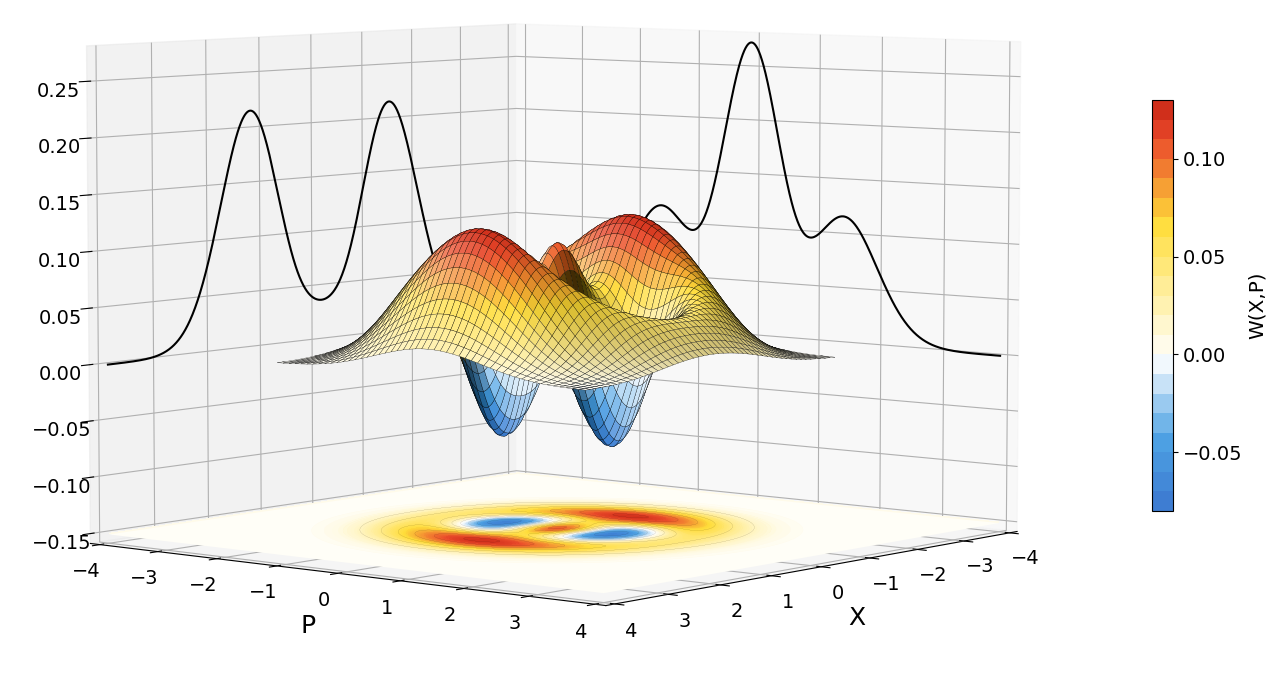} }}
    \caption{\justifying Simulation of the even cat-state generated with the experiment before the storage inside the QMC.}
    \label{fig:simulated_even}
\end{figure}

Our experiment also produces even cat-states when the second heralded state is a single-photon state instead of a $|2\rangle$ state. The generation rate is much higher than the odd cat-state generation. The Figure~\ref{fig:simulated_even} shows the simulated state.

This state has a fidelity of 0.61 with an even cat-state of amplitude $\alpha = 1.71$ squeezed of 3.90 dB.

\subsubsection{Experimental result \label{appendix_even_cat}}
\begin{figure}[h!]
    \centering
    \includegraphics[width=0.8\linewidth]{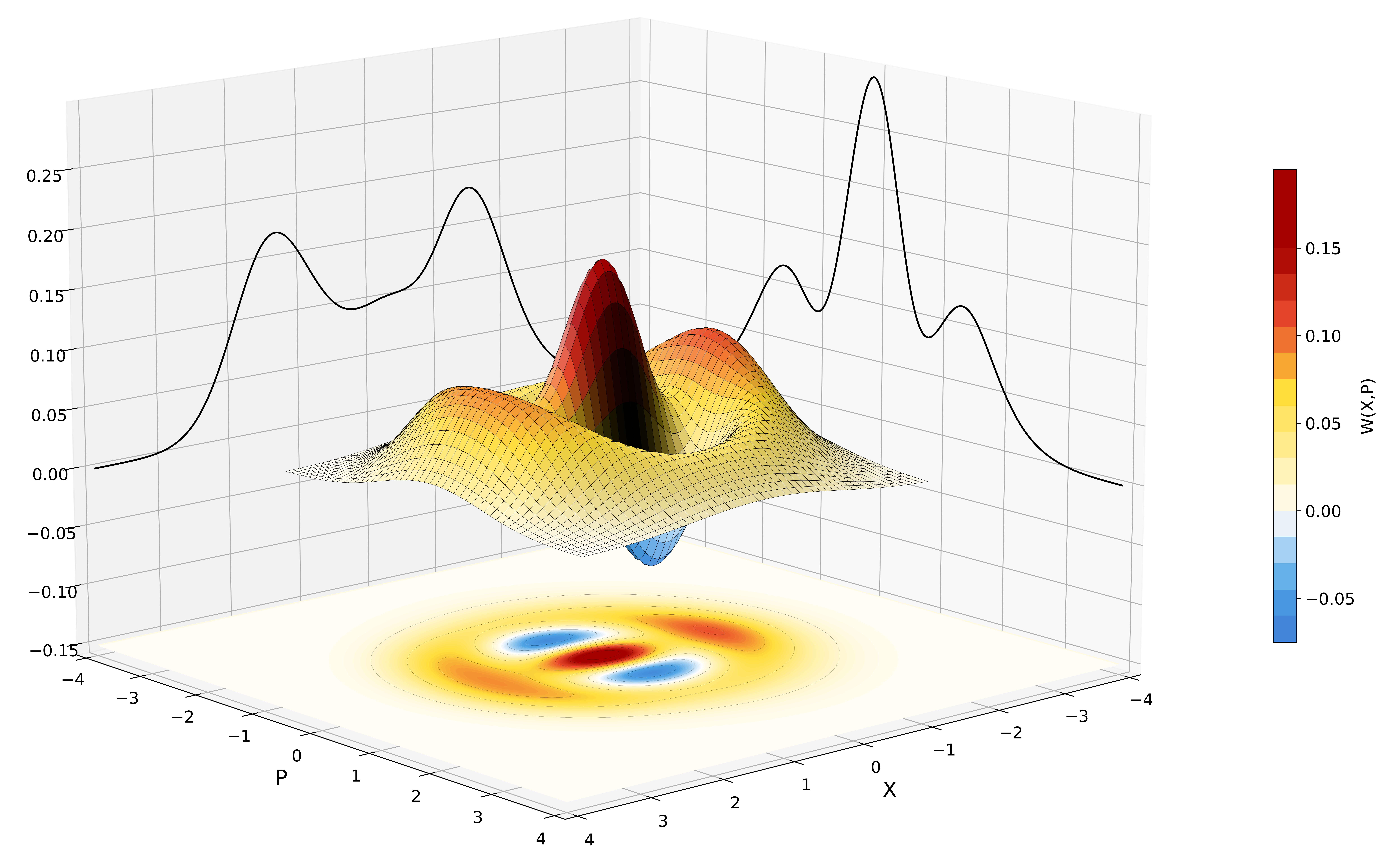}
    \caption{\justifying Reconstructed Wigner function of the generated even cat-states corrected for the detection efficiency and for the storage in the QMC.}
    \label{fig:even_cat}
\end{figure}

A typical even cat-state generated when realizing the coalescence of two single photons in the protocol leads to the Figure~\ref{fig:even_cat}.
The fidelity of the state (corrected from the efficiency of the homodyne detection and the storage) is $F \sim 0.58$ with an even cat-state of amplitude $\alpha = 1.71$ squeezed of 3.90 dB, at generation rate of $\sim 1 $kHz.

\section{Electronic sequence and phase measurement \label{appendix_protocole}}
When a first photon is heralded, a pulse triggers a fast, low-loss Pockels Cell from Raicol company, acting as a half-wave plate. This flips the polarization of the single photon, thereby storing it in the Quantum Memory Cavity. Simultaneously, a waveform generator initiates an 850 ns linearly decreasing voltage ramp. When a second Fock state is heralded, the voltage at that moment is recorded. By defining non-overlapping bins in the resulting voltage histogram, one can determine without ambiguity the number of round trips the first photon has experienced in the cavity. At this stage, the second state can be either $|1\rangle$ or $|2\rangle$. This is why an independent delay generator is triggered only when both APDs click at the same time, sending a pulse to the oscilloscope allowing post-selection of the coincidence events. Consequently, the experiment also produces even cat-states at kHz rate.

Then, applying another arbitrary pulse ensures that both the stored and newly heralded state interact with the PC at the right point of its rise time, when the applied voltage corresponds to the required retardance $\delta$. This is possible thanks to the low jitter ($<$25 ps) of our electronic control system.

At this point, we perform a quadrature measurement. If the result is close to zero, we obtain a cat-state stored within the QMC. Another pulse is then applied to flip the polarization of this stored state. It is important to note that since the wave plate operation is implemented during the rise time of the pulse, the generated state can be directly extracted by slightly extending the pulse duration, allowing the $\lambda/2$ operation to immediately release the stored cat-state. Here, we characterize this state using the same homodyne detection as for the heralding measurement; therefore, we have to store the state because of the limited bandwidth of this homodyne detection.

After 15 round trips in the cavity, the generated states are characterized via quadrature measurements $X_\theta$. To perform quantum state tomography, we need to measure quadratures at different phases.
Assuming the local oscillator’s phase remains stable during a few dozens of $\mu s$, the final quadrature measurement corresponds to the phase accumulated by the stored state. To determine this phase, coherent states are sent into the QMC between cat-state generation cycles, using a Pockels cell in amplitude modulation mode. By measuring interference between the coherent states and the local oscillator, we can retrieve the relative phase.
To extract this phase, the coherent states are periodically $\pi/2$ phase-shifted using a PC modulated at a frequency (5 kHz) much higher than the QMC’s phase noise. By subtracting the phase of a non-stored coherent state from that of a stored one, we can determine the phase shift introduced by the storage process.

As previously noted, the bandwidth of the homodyne detection system (2 MHz) is lower than the repetition rate of the Ti:Sapphire laser (76 MHz). If the local oscillator were a simple pick-off from the Ti:Sapphire pulse train, each homodyne measurement would be contaminated by the preceding pulses, resulting in unacceptable noise from vacuum contributions. To mitigate this, a fast PC is used in amplitude modulation mode to selectively transmit only the relevant pulses for detection.

\section{Characterization of the single photon source and QMC losses \label{appendix_losses}}
\begin{figure}[]
    \centering
    \includegraphics[width=0.95\linewidth]{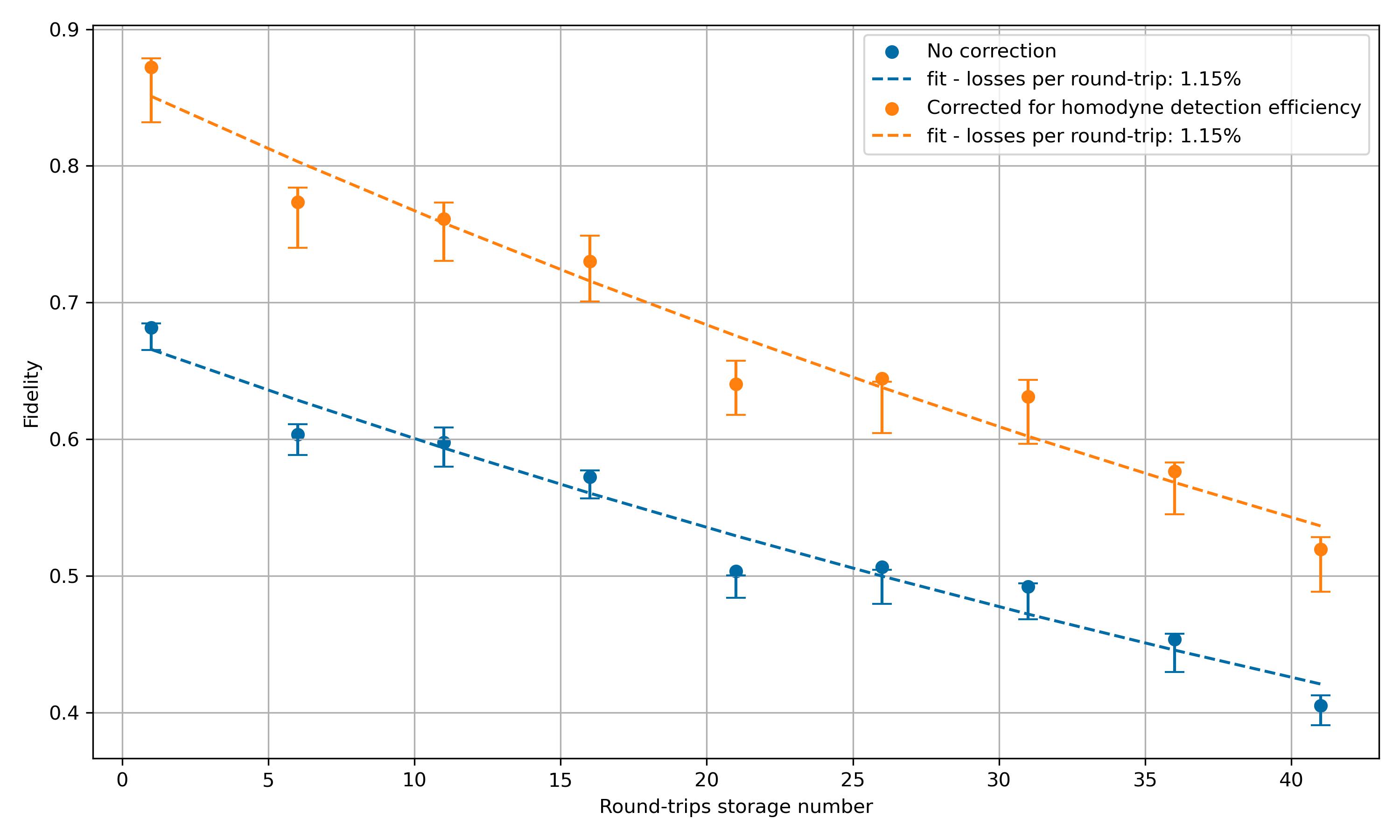}
    \caption{\justifying Fidelity of the single photons with respect to their number of storage round trips, with and without correction for the homodyne detection efficiency. The dotted lines represent the fitted exponential decay. The error bars are obtained from bootstrap resampling.}
    \label{fig:photon}
\end{figure}

To estimate the fidelity of the single photons with respect to their number of storage round trips inside the QMC, we perform the tomography of the photons for different numbers of storage round trips. We extract the losses per round trip from the exponential decay of these fidelities with the number of storage round trips. The results are depicted on Figure~\ref{fig:photon}.

The estimated losses per round trip are $\sim 1.15\%$. Therefore, the reconstructed Wigner functions presented in this article are not overcorrected, since we fixed losses to be $1\%$ per round trip. 

\bibliography{references}

\end{document}